\documentclass[12pt,aps,prd,superscriptaddress,showpacs,floatfix,nofootinbib]{revtex4-1}
\usepackage[utf8]{inputenc}

\usepackage{bmpsize}

\usepackage{hyperref}
\usepackage{color}
\usepackage{graphicx}	% Include figure filed
\usepackage{bm}
\usepackage{amsmath}
\usepackage{amsfonts}
\usepackage{eufrak}

\begin{document}

\title{Azimuthal Anisotropies at High Momentum from Purely Non-Hydrodynamic Transport}

\author{Paul Romatschke}
\affiliation{Department of Physics, University of Colorado, Boulder, Colorado 80309, USA}
\affiliation{Center for Theory of Quantum Matter, University of Colorado, Boulder, Colorado 80309, USA}
%\emailAdd{paul.romatschke@colorado.edu}
%\date{\today}

\begin{abstract}
  In the limit of short mean free path, relativistic kinetic theory gives rise to hydrodynamics through a systematically improvable gradient expansion.   In the present work, a systematically improvable expansion in the opposite limit of large mean free path is considered, describing the dynamics of particles which are almost, but not quite, non-interacting. This non-hydrodynamic ``eremitic'' expansion does not break down for large gradients, and may be useful in situations where a hydrodynamic treatment is not applicable. As applications, azimuthal anisotropies at high transverse momenta in Pb+Pb and p+Pb collisions at $\sqrt{s}=5.02$ TeV are calculated from the first order eremitic expansion of kinetic theory in the relaxation time approximation.
\end{abstract}

\maketitle
%%%%%%%%%%%%%
%%%%%%%%%%%%%
\section{Introduction}
%%%%%%%%%%%%%
%%%%%%%%%%%%%

Is there a simple description for transport when hydrodynamics fails?

At low momenta, relativistic hydrodynamics has been tremendously successful in offering quantitative descriptions and predictions of experimental data from high energy nuclear collisions (see Ref.~\cite{Florkowski:2017olj,Romatschke:2017ejr,Nagle:2018nvi,Busza:2018rrf} for recent reviews.) Hydrodynamics breaks down when non-hydrodynamic modes start to dominate over hydrodynamic modes, and it has been suggested that the experimentally observed peak in azimuthal anisotropies at transverse momenta $p_T\simeq 4$ GeV indicates the transition from hydrodynamic to non-hydrodynamic transport \cite{Romatschke:2016hle}.
For conformal field theories at weak (strong) coupling, this transition happens at a well-defined momentum scale $k_c$ \cite{Romatschke:2015gic,Grozdanov:2016vgg}. For momenta above $k_c$, the lifetimes of hydrodynamic modes are shorter than those from non-hydrodynamic modes, hence at late times bulk transport will be dominated by purely non-hydrodynamic degrees of freedom. While many studies exist that include both hydrodynamic and non-hydrodynamic modes, little is known about the phenomenological implications and observational consequences of \textit{pure} non-hydrodynamic transport where all hydrodynamic mode contributions have been turned off. The present work is meant as a step in this direction by studying non-hydrodynamic transport for the case of relativistic kinetic theory.

The kinetic theory of classical gases has a long history \cite{Boltzmann:1872}, yet active research on its properties is still ongoing. Recent examples include 
%
%As examples, let me mention the emergence of fluid dynamics from kinetic theory in the limit of short mean free path \cite{CE},
the divergence of the gradient expansion of kinetic theory \cite{PhysRevLett.56.1571,Denicol:2016bjh}, its non-perturbative resummation leading to hydrodynamic attractors\cite{Heller:2016rtz,Romatschke:2017vte,Romatschke:2017acs,Florkowski:2017jnz,Denicol:2017lxn,Behtash:2017wqg,Blaizot:2017ucy}, the characterization of the non-hydrodynamic modes in kinetic theory \cite{Romatschke:2015gic,Kurkela:2017xis} and the ``Lattice Boltzmann Approach'' which uses kinetic theory as an efficient algorithm to simulate fluid dynamics \cite{Mendoza:2009gm,Romatschke:2011hm,Succi:2014}.

%In this context it should also be mentioned that the equations of Navier and Stokes, which emerge from the Boltzmann equation in the limit of small gradients, famously have unsolved questions about the existence of their solutions \cite{millenium}, as well as suffering from faster-than-light signal propagation in the relativistic context \cite{Hiscock:1983zz,Hiscock:1985zz}. In view of these challenges, there continues to be ample motivation to study the properties of kinetic theory since it provides a physically motivated completion of the Navier-Stokes equations.

For vanishing mean free path, kinetic theory corresponds ideal (non-viscous) fluid dynamics that is described by the Euler equation \cite{Euler}. Small, but non-vanishing mean-free path corrections give rise to viscous fluid dynamics, described by the equations of Navier and Stokes \cite{Navier:1822,Stokes:1845}. Higher order corrections to the small mean free path regime can be systematically calculated \cite{Burnett}. The opposite limit of large mean free path is known as rarefied gas dynamics or high Knudsen number regime \cite{1909AnP...333..999K,2001ApMRv..54B..90C,2014PhyA..413..409S}, and in the extreme case of infinite mean free path gives rise to non-interacting (or free-streaming) particle dynamics. For infinite mean free path, the classic kinetic equations can be solved analytically using the method of characteristics, leading to ballistic evolution. In a sense, ballistic evolution and ideal fluid dynamics are analogues of each other, corresponding to opposite extreme limits of infinite and zero mean free path, respectively. However, while the systematic small mean free path expansion has been recognized to lead to viscous fluid dynamics, the equivalent systematic expansion at large but finite mean free path seems to have received less attention in the high energy physics literature, except for two works \cite{Heiselberg:1998es,Borghini:2010hy}.  The present work is meant to consider the phenomenological consequences arising from such a systematic expansion, extending in particular the pioneering work by Borghini and Gombeaud \cite{Borghini:2010hy} to the case of large momenta. For large mean free path, particles rarely interact, similar to hermit crabs in their natural environment, hence this systematic expansion will be referred to as ``eremitic'' expansion in the following.

\section{Setup}

I will consider a system of massless on-shell classical particles with a continuum distribution of locations ${\bf x}$ and momenta ${\bf p}$ at any given time $t$. Because the particles are massless, their dynamics will be governed by relativistic kinetic theory, although it should be straightforward to modify the discussion for massive particles with non-relativistic dynamics. The relativistic Boltzmann equation is given by
\begin{equation}
   \label{eq:boltz}
  p^\mu \partial_\mu f(t,{\bf x},{\bf p})= -{\cal C}[f]\,,
\end{equation}
where $f(t,{\bf x},{\bf p})$ is the on-shell phase-space particle distribution function, $p^\mu=\left(p^0,{\bf p}\right)$ is the particle's four momentum, and ${\cal C}[f]$ is the collision kernel which has the property that it vanishes both in equilibrium  as well as for non-interacting particles. The collision kernel depends on the details of the particle interactions, and is usually a (complicated) functional of the particle distribution function $f$. In order to give a more hands-on treatment, it will be useful to consider a concrete and simple example for the collision kernel, such as 
the relaxation time (or BGK \cite{PhysRev.94.511}) approximation where
\begin{equation}
  \label{eq:BGK1}
  {\cal C}[f]=-\frac{p^\mu u_\mu}{\tau_R}\left(f-f_{\rm eq}[f]\right)\,.
\end{equation}
In this equation, $\tau_R$ is the relaxation time (proportional to the mean free path), $u^\mu=\left(u^0,{\bf u}\right)$ is a collective four-velocity vector and $f_{\rm eq}[f]$ is the pseudo-equilibrium distribution function for a configuration given by $f$. For this work, the mostly plus metric convention $g_{\mu\nu}={\rm diag}(-,+,+,+)$ will be used such that $p^\mu u_\mu=-p^0 u^0+{\bf p}\cdot {\bf u}$.

If the system was in equilibrium with a temperature $T$ in a local rest frame given by the four vector $u^\mu$ in some global coordinate system, then the equilibrium distribution function for classical particles can be taken as
\begin{equation}
  \label{eq:feq}
  f_{\rm eq}=\frac{\pi^2}{3}e^{p^\mu u_\mu/T}\,.
\end{equation}
Out of local equilibrium, there strictly speaking is no temperature, but for classical particles, a local rest frame and an energy density can always \cite{Arnold:2014jva} be  obtained from the local energy-momentum tensor \cite{DeGroot:1980dk,Romatschke:2011qp} 
\begin{equation}
  \label{eq:emt}
  T^{\mu\nu}(t,{\bf x})=\int \frac{d^4 p}{(2\pi)^4}2 \theta(p^0) \delta\left(p^\alpha p_\alpha\right) p^\mu p^\nu f\left(t,{\bf x},{\bf p}\right)\,,
\end{equation}
where $\theta(x)$ denotes a step-function. From the energy-momentum tensor, the local energy density $\epsilon(t,{\bf x})$ and local rest-frame four vector $u^\mu(t,{\bf x})$ can be obtained as the time-like eigenvector and associated eigenvalue,
\begin{equation}
  \label{eq:eval}
  u_\mu T^{\mu\nu}=-\epsilon u^\nu\,,
\end{equation}
together with the normalization constraint $u^\mu u_\mu=-1$. For massless particles in local thermodynamic equilibrium where $f=f_{\rm eq}$, the energy density obtained as the time-like eigenvector of $T^{\mu\nu}$ is related to the temperature $T$ as
\begin{equation}
  \epsilon=T^4\,.
\end{equation}
Out of equilibrium, I define a pseudo-temperature (also denoted by $T$) from the energy density as $T=\epsilon^{1/4}$ (cf. the discussion in Ref.~\cite{Romatschke:2017ejr}). This pseudo-temperature, together with the time-like eigenvector $u^\mu$ of $T^{\mu\nu}$, are used to define the pseudo-equilibrium distribution function $f_{\rm eq}[f]$ via Eq.~(\ref{eq:feq}), where the functional dependence on the non-equilibrium particle distribution $f$ has been denoted explicitly.

\subsection{Review of small mean free path expansion}

For small mean free path, the system is close to equilibrium and the collision kernel is almost vanishing, ${\cal C}[f_{\rm eq}]\simeq 0$.
In the relaxation time approximation, the mean free path in kinetic theory is proportional to 
$\tau_R$, cf. Ref.~\cite{Romatschke:2017ejr}. To set the stage, consider first the well-known hydrodynamic gradient expansion of Eq.~(\ref{eq:boltz}) for small mean free path $\tau_R T\ll 1$. To simplify the analytic treatment, I will consider the case of a conformal system where  $\tau_R T ={\rm const}$, but one can expect results to generalize to non-conformal cases.
To leading (zeroth) order in $\tau_R T\ll 1$, Eq.~(\ref{eq:boltz}) leads to
\begin{equation}
  f=f_{\rm fluid, (0)}=f_{\rm eq}\,,
\end{equation}
which gives rise to an energy momentum tensor of the form
\begin{equation}
  T_{{\rm fluid}, (0)}^{\mu\nu}=(\epsilon+P) u^\mu u^\nu+P g^{\mu\nu}\,,
\end{equation}
where $P=\frac{\epsilon}{3}$ is the local equilibrium pressure for a gas of massless particles. Conservation of this energy-momentum tensor $\partial_\mu T^{\mu\nu}=0$ can be recognized as the relativistic equation of continuity and the Euler equation, respectively. Thus the zeroth order expansion in small mean free path corresponds to zeroth order, or ideal, fluid dynamics.

To first order in the small mean free path expansion of Eq.~(\ref{eq:boltz}) one has
\begin{equation}
  \label{eq:hy1}
  f\simeq f_{\rm fluid, (0)}+f_{\rm fluid, (1)}\,,\quad
  {\cal C}[f]\simeq \left.\frac{\delta {\cal C}[f]}{\delta f}\right|_{f=f_{\rm fluid,(0)}}f_{\rm fluid, (1)}
%  f_{\rm fluid, (1)}=\frac{\tau_R}{p^\mu u_\mu} p^\mu \partial_\mu f_{\rm eq}[f_{\rm fluid, (0)}]\,,
\end{equation}
with \cite{Romatschke:2017ejr}
\begin{equation}
  \label{eq:ff1}
  f_{\rm fluid, (1)}=\frac{\tau_R}{p^\lambda u_\lambda} p^\mu \partial_\mu f_{\rm fluid, (0)}=\tau_R f^\prime_{\rm eq} \frac{p^\mu p^\nu \sigma_{\mu\nu}}{2 T p^\lambda u_\lambda}\,,
\end{equation}
in the relaxation time approximation for ${\cal C}$ where $f^\prime_{\rm eq}$ denotes the derivative of the equilibrium distribution function with respect to $p^\mu u_\mu/T$ and $\sigma^{\mu\nu}=\partial^\mu u^\nu+\partial^\nu u^\mu -\frac{2}{3} (g^{\mu\nu}+u^\mu u^\nu) \partial_\lambda u^\lambda$ is the shear-stress tensor.
Calculating the energy-momentum tensor for $f=f_{\rm fluid, (0)}+f_{\rm fluid, (1)}$ one finds
\begin{equation}
  \label{eq:t1}
  T_{\rm fluid, (0)}^{\mu\nu}+T_{\rm fluid, (1)}^{\mu\nu}=(\epsilon+P) u^\mu u^\nu+P g^{\mu\nu}-\eta \sigma^{\mu\nu}\,.
\end{equation}
Here $\eta$ is the shear viscosity coefficient that for a conformal system is usually expressed as a ratio with respect to the pseudo-equilibrium entropy density $s=\frac{\epsilon+P}{T}$. For the kinetic theory at hand, one finds $\frac{\eta}{s}=\frac{\tau_R T}{5}$ \cite{Romatschke:2017ejr}.
With  the energy-momentum tensor given by Eq.~(\ref{eq:t1}), the conservation equations $\partial_\mu T^{\mu\nu}=0$ can be recognized as the relativistic Navier-Stokes equations. Therefore, the first order expansion in small mean free path corresponds to first order, or viscous, fluid dynamics. Higher expansion orders may be systematically generated using this procedure.

Note that in the small mean free path expansion, the relevant expansion parameter is $\tau_R$ times a typical gradient strength, cf. Eq.~(\ref{eq:ff1}). This implies that the expansion fails for large gradients (see the discussion in Ref.~\cite{Romatschke:2015gic}). 

\subsection{Zeroth order eremitic expansion: ballistic regime}

Let me now consider an ``eremitic'' expansion of the kinetic theory in the regime of large mean free path where ${\cal C}[f]\simeq 0$ or $\tau_R T\gg 1$ in the relaxation time approximation. To leading (zeroth) order in eremitic expansion,  Eq.~(\ref{eq:boltz}) leads to
\begin{equation}
  \label{eq:ballistic}
  \left(\partial_t+{\bf v}\cdot \nabla\right) f(t,{\bf x},{\bf p})=0\,, \quad {\bf v}\equiv \frac{\bf p}{p^0}\,.
\end{equation}
This equation can be solved analytically using the method of characteristics:
$$
  \frac{d{\bf x}(t)}{dt}={\bf v}\,,\quad \frac{d f}{d t}=0\,,
$$
  leading to
  \begin{equation}
    f=f_{\rm hermit, (0)}(t,{\bf x},{\bf p})=f_{\rm init}\left({\bf x}-{\bf v} t,{\bf p}\right)\,,
  \end{equation}
  where $f_{\rm init}({\bf x},{\bf p})$ is the particle distribution function at initial time $t=0$. The energy-momentum tensor for the zeroth order eremitic expansion is given by (\ref{eq:emt}), which requires specification of $f_{\rm init}$. Some example cases will be considered below. 

\subsection{First order eremitic expansion}

The first order eremitic expansion for the distribution function is given by
\begin{equation}
f=  f_{\rm hermit, (0)}+f_{\rm hermit, (1)}\,,\quad p^\mu \partial_\mu f_{\rm hermit, (1)}= -{\cal C}[f_{\rm hermit, (0)}]\,,
\end{equation}
with
\begin{equation}
  p^\mu \partial_\mu f_{\rm hermit, (1)}= \frac{p^\mu u_\mu}{\tau_R}\left(f_{\rm hermit, (0)}-f_{\rm eq}[f_{\rm hermit, (0)}]\right)\,,
  \end{equation}
    in the relaxation time approximation where the free-streaming result (\ref{eq:ballistic}) has been used. The defining equation for $f_{\rm hermit, (1)}$ is similar to (\ref{eq:ballistic}), but with a non-vanishing constant on the rhs. Using again the method of characteristics, one finds for the first order eremitic correction
\begin{equation}
  \label{eq:hermit1}
  f_{\rm hermit, (1)}(t,{\bf x},{\bf p})=\int_0^tdt^\prime\left. \frac{p^\mu u_\mu(t^\prime, {\bf x}^\prime)}{\tau_R p^0}\left(f_{\rm hermit, (0)}(t^\prime, {\bf x}^\prime,{\bf p})-f_{\rm eq}\left[f_{\rm hermit, (0)}(t^\prime,{\bf x}^\prime,{\bf p})\right]\right)\right|_{{\bf x}^\prime={\bf x}-{\bf v} t}
  \end{equation}

Higher order eremitic corrections can be obtained systematically by repeating this procedure. Note that in the large mean free path expansion, the relevant expansion parameter is $\tau_R^{-1}$ times an integral over the characteristic, cf. Eq.~(\ref{eq:hermit1}). This implies that the eremitic expansion fails for late times/large distances, which corresponds to the case of small gradients. This is because the expansion is around non-interacting particles, and corrections from interactions build up coherently along the characteristic for small gradients.

\subsection{Collective Modes}

\begin{figure}[t]
  \includegraphics[width=\linewidth]{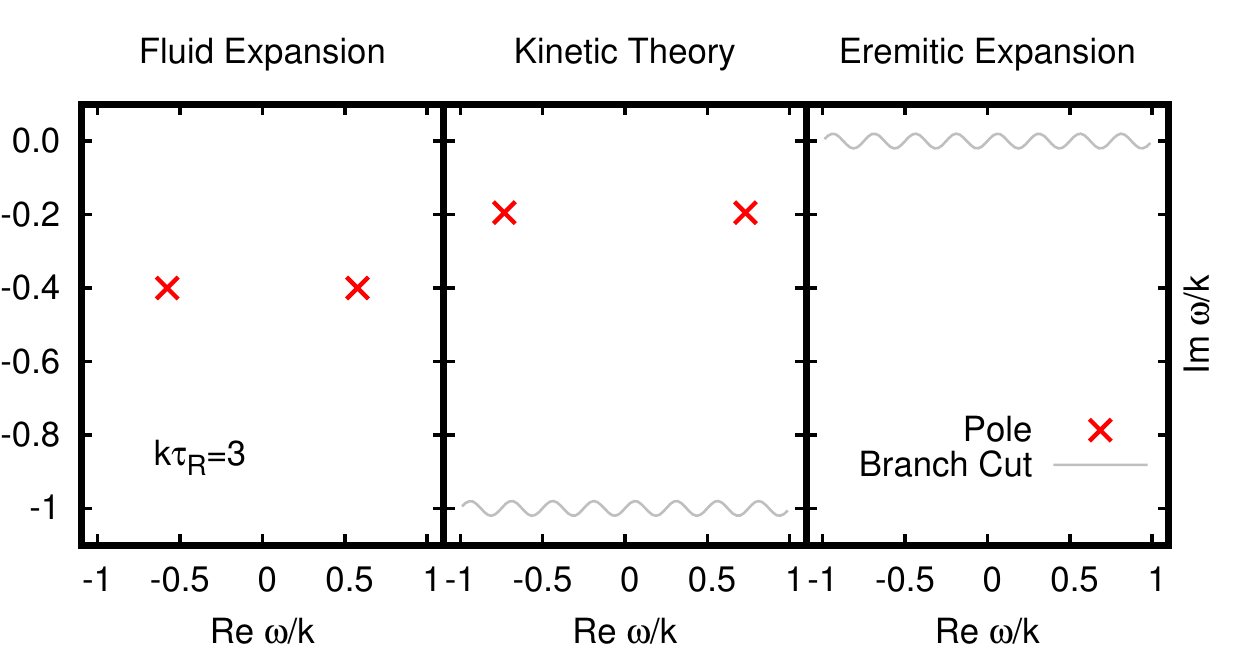}
  \caption{\label{fig:one} Collective sound mode structure in the complex frequency plane for kinetic theory in the relaxation time approximation (middle panel), small mean free path (fluid) expansion (left panel) and large mean free path (hermit) expansion (right panel). Simple poles of the retarded correlator are shown as crosses, while logarithmic branch cuts are shown as lines.}
\end{figure}

Because hydrodynamic and eremitic expansions are opposite limits of kinetic theory (\ref{eq:boltz}), their collective mode structure can also be expected to be different. The collective modes of Eq.~(\ref{eq:boltz}) in the relaxation time approximation have been analyzed in Ref.~\cite{Romatschke:2015gic} for constant $\tau_R T$ (see Ref.~\cite{Kurkela:2017xis} for the case of momentum dependent relaxation time), and results are summarized here in order to keep this work self-contained.

The collective modes can be calculated as the singularities of the retarded two-point function $G_R(\omega,{\bf k})$ of the energy-momentum tensor, with $\omega,{\bf k}$ the conjugate Fourier momenta to $t,{\bf x}$ (see e.g. Ref.~\cite{Romatschke:2017ejr}). For simplicity, I will only discuss the singularities of $G_R$ in sound channel. For fluid dynamics, the singularities of $G_R(\omega,{\bf k})$ in the sound channel are simple poles located at 
\begin{equation}
  \label{eq:hycm}
  \omega_{\rm fluid, (1)}^{\pm}=\pm c_s |{\bf k}|-i \frac{2 \tau_R {\bf k}^2}{15}\,.
  %=\pm c_s |{\bf k}|-i \frac{2 \eta {\bf k}^2}{3 s T}\,.
\end{equation}
These are the familiar sound modes.

By contrast, the collective modes for the eremitic expansion can directly be obtained by a Fourier transform of $p^\mu \partial_\mu f_{\rm hermit, (1)}$ using the setup in Ref.~\cite{Romatschke:2015gic}. One finds that the collective modes in the eremitic expansion are logarithmic branch cuts emanating from the branch points
\begin{equation}
  \label{eq:ermit}
  \omega_{\rm hermit, (1)}^{\pm}=\pm |{\bf k}|\,.
  \end{equation}

Not surprisingly, the analysis of the collective modes contained in (\ref{eq:boltz}) for general $\tau_R$ contains both of these types of singularities, logarithmic cuts emanating from branch points 
\begin{equation}
  \omega_{\rm cut}^{\pm}=\pm |{\bf k}|-\frac{i}{\tau_R}\,,
\end{equation}
and hydrodynamic poles located at $\omega_{\rm hydro}^\pm(k)$\footnote{For a common choice of the location of the logarithmic branch cut, the hydrodynamic poles move through the cut onto the next Riemann sheet for $k \tau_R\gtrsim 4.5313912\dots\equiv k_c \tau_R$ and thus are no longer present on the fundamental Riemann sheet \cite{Romatschke:2015gic}. However, this behavior is not generic because other choices of the logarithmic cut location may be employed \cite{Kurkela:2017xis}. What is generic, however, is that for $k>k_c$ the hydrodynamic poles are farther away from the real axis then the non-hydrodynamic branch cut, implying late-time transport to be dominated by non-hydrodynamic degrees of freedom.}. The hydrodynamic poles 
approach the fluid dynamic results (\ref{eq:hycm}) in the limit of small $\tau_R |{\bf k}|$, as they should. The branch points approach the eremitic results (\ref{eq:ermit}) in the limit of large $\tau_R |{\bf k}|$, as they should. The situation is summarized in Fig.~\ref{fig:one}, which depicts the singularity structure of the sound channel two-point function in the complex frequency plane.

It is common to refer to the collective modes corresponding to the sound poles $\omega_{\rm hydro}$ as hydrodynamic modes, and label the modes corresponding to the branch cuts $\omega_{\rm cut}$ as non-hydrodynamic modes. Thus, the fluid dynamic expansion of kinetic theory contains only hydrodynamic modes, the eremitic expansion contains only non-hydrodynamic modes, while kinetic theory without any expansion contains both.

\section{Analytic Examples}

While powerful methods exist to solve the Boltzmann equation (\ref{eq:boltz}) numerically, the strength of the eremitic expansion lies in the possibility of obtaining analytic (or at least semi-analytic) results. To this end, let me point out some examples where analytic treatments are possible. All of these examples will be within a class of initial particle distribution functions that can be written as
\begin{equation}
  f_{\rm init}({\bf x},{\bf p})=F(|{\bf p}|/\Lambda({\bf x}))\,,
\end{equation}
with $F$ and $\Lambda$ arbitrary functions. Within this class of examples, the zeroth order eremitic expansion leads to
\begin{equation}
   f_{\rm hermit, (0)}(t, {\bf x},{\bf p})=F(p/\Lambda({\bf x}-{\bf v}t))\,,\quad p\equiv |{\bf p}|\,,
\end{equation}
and the associated energy momentum tensor is
\begin{equation}
T^{\mu\nu}_{\rm hermit, (0)}=\int \frac{d\Omega}{4 \pi} v^\mu v^\nu \Lambda^4\left({\bf x}-{\bf v}t\right) \int \frac{d p}{2 \pi^2} p^3 F(p)\,, \quad v^\mu\equiv \left(1,{\bf v}\right)\,.
\end{equation}
Examples of $\Lambda,F$ where $T^{\mu\nu}_{\rm hermit, (0)}$ can be calculated analytically will be discussed in the following.

\subsection{Single Gaussian Hot-Spot}

\begin{figure}[t]
  \includegraphics[width=0.5\linewidth]{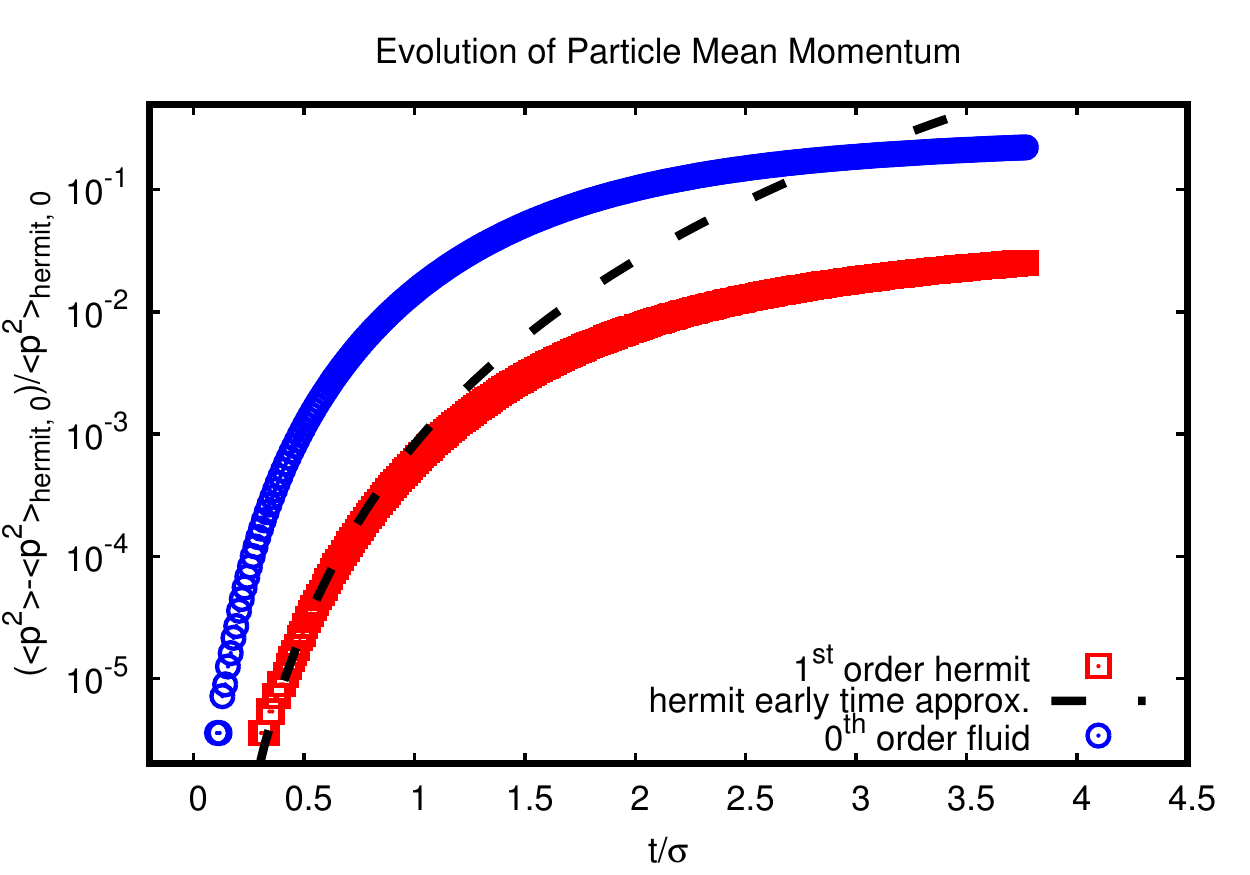}\hfill
  \includegraphics[width=0.5\linewidth]{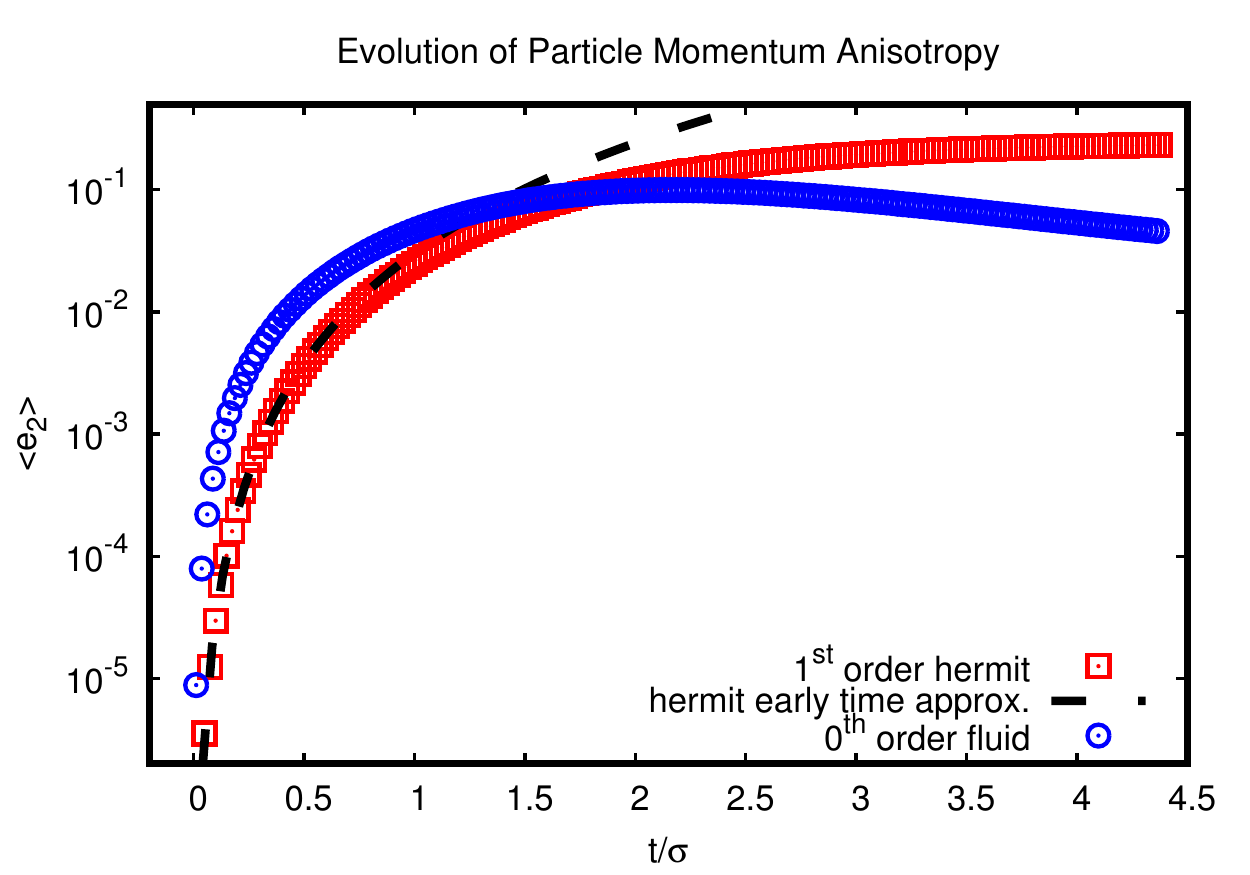}
  \caption{\label{fig:two} Left: Time evolution of particle mean square momentum $\langle p^2\rangle$ for a single Gaussian hot-spot of width $\sigma$. Right: Time evolution of mean elliptic momentum anisotropy for two Gaussian hot-spots located at ${\bf x}=\mp \sigma$. Show are numerical results for first order eremitic expansion and early-time approximations thereof for $\tau_R T=1$, as well as numerical results for $0^{\rm th}$ order fluid expansion.}
\end{figure}

The first example is that of a single Gaussian hot-spot, with an initial pseudo-temperature distribution given by
\begin{equation}
  \Lambda({\bf x})=T_{\rm init} e^{-\frac{\left({\bf x}-{\bf x}_0\right)^2}{8 \sigma^2}}\,,
\end{equation}
with $\sigma,{\bf x}_0,T_{\rm init}$ controlling the width, location, and height of the hot-spot, respectively. In addition, let me consider the case where the initial particle distribution function is in local equilibrium,
\begin{equation}
  \label{eq:localeq}
  f_{\rm init}({\bf x},{\bf p})=\frac{\pi^2}{3}e^{-p/\Lambda({\bf x})}\,.
\end{equation}
For this case, the zeroth order eremitic expansion leads to
\begin{equation}
  f_{\rm hermit, (0)}=\frac{\pi^2}{3}e^{-p/\Lambda({\bf x}-{\bf v}t)}\,,
\end{equation}
with energy momentum tensor components given by
\begin{eqnarray}
  \label{eq:emt0}
  T^{00}_{\rm hermit, (0)}&=&T_{\rm init}^4 e^{-\frac{({\bf x}-{\bf x}_0)^2+t^2}{2 \sigma^2}}  \frac{ \sinh a}{a}\,,\nonumber\\
  T^{0i}_{\rm hermit, (0)}&=& T_{\rm init}^4 e^{-\frac{({\bf x}-{\bf x}_0)^2+t^2}{2 \sigma^2}} \frac{x^i-x_0^i}{|{\bf x}-{\bf x}_0|} \left(\frac{\cosh a}{a}-\frac{\sinh a}{a^2}\right)\,,\\
  T^{ij}_{\rm hermit, (0)}&=&T_{\rm init}^4 e^{-\frac{({\bf x}-{\bf x}_0)^2+t^2}{2 \sigma^2}}\delta^{ij} \left(\frac{ \cosh a}{a^2}-\frac{ \sinh a}{a^3}\right)\nonumber\\
  &&+T_{\rm init}^4 e^{-\frac{({\bf x}-{\bf x}_0)^2+t^2}{2 \sigma^2}}\frac{\left(x^i-x_0^i\right)\left(x^j-x_0^j\right)}{|{\bf x}-{\bf x}_0|^2}
  \left(\frac{ \sinh a}{a^3}(3+a^2)-\frac{3\cosh a}{a^2}\right)\,,\nonumber
\end{eqnarray}
where $a \equiv \frac{t |{\bf x}-{\bf x}_0|}{\sigma^2}$. The pseudo-temperature $T$ and flow vector $u^\mu$ corresponding to this energy momentum tensor can be obtained via eigenvalue decomposition, cf. Eq.~(\ref{eq:eval}). One finds
\begin{eqnarray}
  \label{eq:eigenvals1}
  \epsilon(t,{\bf x})&=&T_{\rm init}^4 e^{-\frac{({\bf x}-{\bf x}_0)^2+t^2}{2 \sigma^2}} \left[\frac{\cosh a}{a^2}-\frac{\sinh a}{a^3}+\sqrt{\left(\frac{\cosh a}{a^2}-\frac{\sinh a}{a^3}\right)^2+\frac{\sinh^2 a}{a^4}-\frac{1}{a^2}}\right]\,,\nonumber\\
  \frac{\bf u}{u^0}&=&\frac{{\bf x}-{\bf x}_0}{|{\bf x}-{\bf x}_0|} \frac{\frac{\cosh a}{a}-\frac{\sinh a}{a^2}}{\frac{\sinh a}{a^3}(1+a^2)-\frac{\cosh{a}}{a^2}+\sqrt{\left(\frac{\cosh a}{a^2}-\frac{\sinh a}{a^3}\right)^2+\frac{\sinh^2 a}{a^4}-\frac{1}{a^2}}}\,,
  \end{eqnarray}
which define $f_{\rm eq}[f_{\rm hermit, (0)}]$ in Eq.~(\ref{eq:hermit1}). For further use I will introduce
\begin{equation}
  T_1(t,|{\bf x}-{\bf x}_0|)=T_{\rm init} e^{-\frac{({\bf x}-{\bf x}_0)^2+t^2}{8 \sigma^2}}\,.
\end{equation}
It is useful to consider the early time regime $t\ll \sigma$ of this case, which can be treated analytically. For small $t$, the eigenvalue decomposition of $T^{\mu\nu}_{\rm hermit, (0)}$ leads to
\begin{eqnarray}
    T(t,{\bf x})&=&T_1(t,|{\bf x}-{\bf x}_0|)\left(1+\frac{t^2 ({\bf x}-{\bf x}_0)^2}{48 \sigma^4}+\ldots\right)\,,\\
    \frac{{\bf u}}{u^0}&=&\left({\bf x}-{\bf x}_0\right)\left(\frac{t}{4 \sigma^2}-\frac{3t^3|{\bf x}-{\bf x}_0|^2 }{320\sigma^6}+\ldots\right)\,,
  \end{eqnarray}
and hence one finds
\begin{equation}
  f_{\rm hermit, (1)}(t, {\bf x},{\bf p})=- f_{\rm init}({\bf x},{\bf p})\frac{t^3 p }{288 \sigma^4 \tau_R T}\left[({\bf x}-{\bf x}_0)^2-3 \left(({\bf x}-{\bf x}_0)\cdot {\bf v}\right)^2\right]+{\cal O}(t^4)\,.
\end{equation}
However, one notes that to this order in $t\ll \sigma$, the space average $\int d^3x f_{\rm hermit, (1)}$ vanishes. The space average of the first order eremitic expansion is given by
\begin{equation}
  \label{eq:firstmp}
  \int d^3x f_{\rm hermit, (1)}(t,{\bf x},{\bf p})=\int_0^t dt^\prime \int d^3x \frac{p^\mu u_\mu(t^\prime, {\bf x})}{p^0 \tau_R}
  \left(f_{\rm hermit, (0)}(t^\prime,{\bf x},{\bf p})-f_{\rm eq}[f_{\rm hermit, (0)}(t^\prime,{\bf x},{\bf p})]\right)\,,
\end{equation}
where the integration variable has been shifted. With the results for $T,u^\mu$ from (\ref{eq:eigenvals1}) one finds
\begin{equation}
  f_{\rm hermit, (0)}-f_{\rm eq}[f_{\rm hermit, (0)}]=\frac{\pi^2}{3}\left[e^{-\frac{p^0}{T_1(t,|{\bf x}-{\bf x}_0|)} e^{-\frac{{\bf v}\cdot ({\bf x}-{\bf x}_0) a}{4 |{\bf x}-{\bf x}_0|}}}-e^{-\frac{p^0}{T_1(t,|{\bf x}-{\bf x}_0|)} u^0\left(1-\frac{{\bf v}\cdot ({\bf x}-{\bf x}_0) a}{4 |{\bf x}-{\bf x}_0|}\right)}\right]\,,
  \end{equation}
where $u^0$ may be obtained from Eq.~(\ref{eq:eigenvals1}) with $u^\mu u_\mu=-1$. 
For early times $t\ll \sigma$ one finds
\begin{equation}
\int d^3x f_{\rm hermit, (1)}(t,{\bf x},{\bf p})\simeq\frac{4 \pi t^5}{5 \sigma^8 \tau_R T}\int d|{\bf x}| |{\bf x}|^6 f_{\rm init}({\bf x},{\bf p})\left(\frac{p}{512 }
-\frac{7 p^2}{7680 \Lambda({\bf x})}+\frac{p^3}{11520 \Lambda^2({\bf x})}\right)\,.
\end{equation}
%As a consequence, observables such as the quadrupole moment $\int d^3 x d^3p p^i p^j f$ will differ at order ${\cal O}(t^3)$ at early times from the free-streaming result.
The mean square particle momentum, defined as
\begin{equation}
  \langle p^2 \rangle[f] = \frac{\int d^3x d^3p p^2 f }
          {\int d^3x d^3p f}\,,
\end{equation}
remains constant as a function of time in the absence of interactions, as can readily be observed from the free-streaming result
\begin{equation}
  \langle p^2 \rangle[f_{\rm hermit, (0)}]=\langle p^2 \rangle[f_{\rm init}]=12 T_{\rm init}^2 \left(\frac{3}{5}\right)^{3/2}\,.
\end{equation}
Interactions can be expected to lead to an increase of $\langle p^2 \rangle$ as a function of time, such that a potential experimental observation of $\langle p^2 \rangle$ carries indirect information about the interaction strength. The time-dependence of $\langle p^2 \rangle$ in first-order eremitic expansion can be found as
\begin{equation}
  \langle p^2 \rangle[f_{\rm hermit, (0)}+f_{\rm hermit, (1)}]\simeq 12 T_{\rm init}^2 \left(\frac{3}{5}\right)^{3/2}\left(1+0.000406634\frac{t^5 T_{\rm init}}{\sigma^4 \tau_R T }+{\cal O}(t^6)\right)\,,
\end{equation}
which  increases as a function of time as expected. Fig.~\ref{fig:two} shows a comparison of this analytic early time approximation to the first order result $\langle p^2 \rangle[f_{\rm hermit, (0)}+f_{\rm hermit, (1)}]$ obtained via numerical integration\footnote{Numerical algorithms employed for this work are publicly available for download from \cite{codedown}.} of Eq.~\ref{eq:firstmp}, confirming the accuracy of the approximation for $t \ll \sigma$. As comparison, I also show the corresponding evolution of
\begin{equation}
  \langle p^2\rangle[f_{\rm fluid, (0)}]=\frac{12 \int d^3x T^5 u^0(2 u_0^2-1)}{\int d^3x T^3 u^0}\,.
  \end{equation}
in Fig.~\ref{fig:two}, which is calculated from (\ref{eq:feq}) using (\ref{eq:eigenvals1}) for the fluid temperature and velocity\footnote{Note that Eq.~(\ref{eq:eigenvals1}) are \textit{not} the solution to ideal fluid equations of motion $\partial_\mu T^{\mu\nu}_{\rm fluid,(0)}=0$ because $\sigma^{\mu\nu}\neq 0$, but they are very close, cf. the discussion in Refs.~\cite{Hatta:2015kia,Romatschke:2015dha}.}

\subsection{Two Gaussian Hot-Spots}
\label{sec:two}

It is possible to generalize the above example to the case of two Gaussian hot-spots. Specifically, consider an initial pseudo-temperature distribution given by
\begin{equation}
  \Lambda({\bf x})=T_{\rm init}\left(e^{-\frac{({\bf x}-{\bf x}_0)^2}{2 \sigma^2}}+ e^{-\frac{({\bf x}-{\bf x}_1)^2}{2 \sigma^2}} \right)^{1/4}
  \end{equation}
with $\sigma,T_{\rm init}$ again controlling the width, and height of the hot-spots, and ${\bf x}_0,{\bf x}_1$ specifying the hot-spot locations, respectively. Taking the distribution at the initial time to be of the form (\ref{eq:localeq}), the zeroth order eremitic energy-momentum tensor $T^{\mu\nu}_{\rm hermit, (0)}$ is given by a simple superposition of (\ref{eq:emt0}) at positions ${\bf x}_0$, ${\bf x}_1$, respectively.
For early times, the eigenvalue decomposition of $T^{\mu\nu}_{\rm hermit,(0)}$ leads to
\begin{eqnarray}
  T(t,{\bf x})&\simeq &\Lambda({\bf x})\left(1+\frac{t^2}{48 \sigma^4}\frac{\left(({\bf x}-{\bf x}_0){\rm e}^0-({\bf x}-{\bf x}_1){\rm e}^1\right)^2+2 {\rm e}^{0+1}\left(({\bf x}-{\bf x}_0)^2+({\bf x}-{\bf x}_1)^2\right)}{\left({\rm e}^0+{\rm e}^1\right)^2}\right)\,,\nonumber\\
  \frac{{\bf u}}{u^0}&\simeq &\frac{t}{4 \sigma^2}\frac{({\bf x}-{\bf x}_0){\rm e}^0+({\bf x}-{\bf x}_1){\rm e}^1}{{\rm e}^0+{\rm e}^1}\,,
\end{eqnarray}
%where
%\begin{equation}
%  T_1(t,{\bf x})=T_{\rm max}\left(e^{-\frac{({\bf x}-{\bf x}_0)^2+t^2}{2\sigma^2}}+e^{-\frac{({\bf x}-{\bf x}_1)^2+t^2}{2\sigma^2}}\right)^{1/4}\,,
%\end{equation}
where the shorthand notations ${\rm e}^{0}=e^{-\frac{({\bf x}-{\bf x}_0)^2}{2\sigma^2}},{\rm e}^{1}=e^{-\frac{({\bf x}-{\bf x}_1)^2}{2\sigma^2}}$ have been used. This leads to
\begin{eqnarray}
  f_{\rm hermit, (1)}&\simeq&-f_{\rm init} \frac{t^3 p}{288 \sigma^4 \tau_R T}\left[
    5\frac{\left(({\bf x}-{\bf x}_0){\rm e}^0+({\bf x}-{\bf x}_1){\rm e}^1\right)^2-3 \left(({\bf x}-{\bf x}_0)\cdot {\bf v} {\rm e}^0+({\bf x}-{\bf x}_1)\cdot {\bf v} {\rm e}^1\right)^2}{\left({\rm e}^0+{\rm e}^1\right)^2}\right.\nonumber\\
    &&\left.-4 \frac{({\bf x}-{\bf x}_0)^2{\rm e}^0+({\bf x}-{\bf x}_1)^2{\rm e}^1-3 \left(({\bf x}-{\bf x}_0)\cdot {\bf v}\right)^2{\rm e}^0-3 \left(({\bf x}-{\bf x}_1)\cdot {\bf v}\right)^2 {\rm e}^1}{{\rm e}^0+{\rm e}^1}\right]\,.
\end{eqnarray}
Without loss of generality, one may take the two Gaussian hot-spots be located at ${\bf x}_0=\left(x_0,0,0\right)$ and ${\bf x}_1=\left(-x_0,0,0\right)$, respectively. An interesting quantity to consider is the evolution of the elliptic momentum anisotropy $e_2$ of the system defined as
\begin{equation}
  \langle e_2\rangle[f]\equiv \frac{\int d^3x \left(T^{yy}-T^{xx}\right)}{\int d^3x \left(T^{yy}+T^{xx}\right)}=\frac{\int d^3x d^3p\, p (v_y^2-v_x^2) f}{\int d^3x d^3p\, p (v_y^2+v_x^2) f}\,,
  \end{equation}
which in the case of $x_0=\sigma$ for early times evaluates to
\begin{equation}
  \langle e_2\rangle[f_{\rm hermit, (0)}+f_{\rm hermit, (1)}]\simeq 0.0149 \frac{t^3 T_{\rm init}}{\sigma^2 \tau_R T}\,.
\end{equation}
(By changing the orders of integration over ${\bf x},{\bf p}$, it is easy to verify that $\langle e_2\rangle [f_{\rm hermit, (0)}]=0$ for all times, cf. Ref.~\cite{Kolb:2000sd}.) A comparison of this result to the numerical evaluation of the first order eremitic expansion is shown in Fig.~\ref{fig:two}, along with the result for the zeroth order hydrodynamic expansion $\langle e_2\rangle[f_{\rm eq}]$, demonstrating that the analytic approximation works well for $t \ll \sigma$.

The present example is very similar to the case of a deformed Gaussian in two dimensions considered in Ref.~\cite{Borghini:2010hy}, where many analytic results were obtained. The main difference to Ref.~\cite{Borghini:2010hy} is that by superposition, the above calculation can be generalized to the case of multiple Gaussian hot-spots without additional complications.

\section{Numerical Examples with Boost Invariance}

Consider now eremitic expansions for particles in a system having boost invariance, as is approximately the case for the high density region of relativistic nucleus-nucleus collisions. To this end, it is useful to consider a coordinate transformation to Milne coordinates proper time $\tau=\sqrt{t^2-z^2}$ and space-time rapidity $\xi={\rm arctanh}\frac{z}{t}$. Using ${\bf x}_T=(x,y)$, the Boltzmann equation in coordinates $x^a=\left(\tau,{\bf x}_T,\xi\right)$ becomes
\begin{equation}
  p^a \partial_a f - \frac{2 p^\xi p^\tau(\tau)}{\tau}\frac{\partial f}{\partial p^\xi}=-{\cal C}[f]\,,
\end{equation}
where $p^\tau(\tau)=\sqrt{p_T^2+(\tau^2 p^\xi)^2/\tau^2}$.
Assuming that the system is invariant under boosts in the longitudinal direction leads to $f=f\left(\tau,{\bf x}_T,{\bf p}_T,p^\xi\right)$, i.e. independent of rapidity. Solution of the characteristic equations for the eremitic expansion to zeroth order gives \cite{Romatschke:2015dha}
\begin{equation}
  f=f_{\rm hermit, (0)}(\tau,{\bf x}_T,{\bf p}_T,p^\xi)=f_{\rm init}\left({\bf x}_T-\frac{\tau {\bf p}_T p^\tau(\tau)}{p_T^2},{\bf p}_T,\tau^2 p^\xi\right)\,,
\end{equation}
and the first order result is given by
\begin{equation}
  \label{eq:firsthermitbjor}
  f_{\rm hermit, (1)}(\tau,{\bf x}_T,{\bf p}_T,p^\xi)=- \int_{\tau_0}^\tau d\tau^\prime\left. \frac{{\cal C}[f_{\rm hermit,(0)}(\tau^\prime, {\bf x}^\prime_T,{\bf p}_T, \tau^2 p^{\xi})]}{p^\tau(\tau^\prime)}\right|_{{\bf x}^\prime_T={\bf x}_T-\frac{{\bf p}_T \left(\tau p^\tau(\tau)-\tau_0 p^\tau_0(\tau)\right)}{p_T^2}}\,,
\end{equation}
where $p^\tau_0(\tau)=\sqrt{p_T^2+(\tau^2 p^\xi)^2/\tau_0^2}$ and integration was started at some finite proper time $\tau_0$. Let us again consider a class of boost-invariant initial particle distribution functions at proper time $\tau=\tau_0$ parametrized by $f_{\rm init}({\bf x}_T,{\bf p}_T,p^\xi)=F\left(p_0^\tau(\tau)/\Lambda\left({\bf x}_T\right)\right)$, such that the associated energy-momentum tensor in zeroth order eremitic expansion is given by
\begin{eqnarray}
  T^{ab}_{\rm hermit, (0)}&=&\int \frac{d^2 p_T dp^\xi \tau}{(2 \pi)^3}\frac{p^a p^b}{p^\tau(\tau)} f_{\rm hermit, (0)}\,,\\
  &=&\int \frac{d\phi dY}{4 \pi} v^a v^b \frac{\Lambda^4\left({\bf x}_T-{\bf v}_T \left(\tau \cosh Y-\tau_0 \sqrt{1+\tau^2/\tau_0^2 \sinh^2{(Y-\xi)}}\right)\right)}{\left(1+\tau^2/\tau_0^2 \sinh^2{(Y-\xi)}\right)^2}  \int \frac{d p_T}{2 \pi^2} p_T^3 F(p_T)\nonumber\,,
  \end{eqnarray}
where I changed variables from $p^\xi=\tau^{-1}p_T \sinh(Y-\xi)$ to momentum rapidity $Y$ and used shorthand notation $v^a=p^a/p_T=\left(\cosh (Y-\xi),{\bf v}_T,\tau^{-1}\sinh(Y-\xi)\right)$. 

\subsection{Single Gaussian Hot-Spot with Boost Invariance}

Taking the initial particle distribution function to be of equilibrium form $F(x)=\frac{\pi^2 Z}{3} e^{-x}$ with $Z$ parametrizing the number of degrees of freedom, and setting $\Lambda({\bf x}_T)=T_{\rm init}e^{-{\bf x}_T^2/(8 \sigma^2)}$ to be a two-dimensional Gaussian leads to simple integral expressions for the non-vanishing components of the energy momentum tensor:
\begin{eqnarray}
  T^{00}_{\rm hermit, (0)}&=&Z T_{\rm init}^4\int_0^\infty dY \frac{\cosh^2 Y e^{-\frac{{\bf x}_T^2+S^2}{2 \sigma^2}}}{\left(1+\tau^2/\tau_0^2 \sinh^2 Y\right)^2} I_0\left(\frac{|{\bf x}_T| S}{\sigma^2}\right)\,,\nonumber\\
  T^{0l}_{\rm hermit, (0)}&=&Z T_{\rm init}^4 \frac{x_T^l}{|{\bf x}_T|} \int_0^\infty dY \frac{\cosh Y e^{-\frac{{\bf x}_T^2+S^2}{2 \sigma^2}}}{\left(1+\tau^2/\tau_0^2 \sinh^2 Y\right)^2} I_1\left(\frac{|{\bf x}_T| S}{\sigma^2}\right)\,,\nonumber\\
  T^{lm}_{\rm hermit, (0)}&=&\frac{Z T_{\rm init}^4 \delta^{lm}}{2} \int_0^\infty dY \frac{ e^{-\frac{{\bf x}_T^2+S^2}{2 \sigma^2}}}{\left(1+\tau^2/\tau_0^2 \sinh^2 Y\right)^2}
 \left[I_0\left(\frac{|{\bf x}_T| S}{\sigma^2}\right)-I_2\left(\frac{|{\bf x}_T| S}{\sigma^2}\right)\right]\nonumber\\
 &&+Z \frac{x_T^l x_T^m}{|{\bf x}_T|} \int_0^\infty dY \frac{ e^{-\frac{{\bf x}_T^2+S^2}{2 \sigma^2}}}{\left(1+\tau^2/\tau_0^2 \sinh^2 Y\right)^2}I_2\left(\frac{|{\bf x}_T| S}{\sigma^2}\right)\,,\nonumber\\
 T^{\xi\xi}_{\rm hermit, (0)}&=&\frac{Z T_{\rm init}^4}{\tau^2}  \int_0^\infty dY \frac{\sinh^2 Y e^{-\frac{{\bf x}_T^2+S^2}{2 \sigma^2}}}{\left(1+\tau^2/\tau_0^2 \sinh^2 Y\right)^2} I_0\left(\frac{|{\bf x}_T| S}{\sigma^2}\right)\,,
 \label{eq:emtmany}
\end{eqnarray}
where $S\equiv \tau \cosh Y- \sqrt{\tau_0^2+\tau^2 \sinh^2 Y}$, $I_n(x)$ denote modified Bessel functions and $l,m=(x,y)$. Given the energy-momentum tensor, one may use (\ref{eq:eval}) to calculate the local energy-density $\epsilon$ and flow vector $u^a=\left(u^\tau, {\bf u}_T,0\right)$ in zeroth order eremitic expansion. These in turn determine ${\cal C}[f_{\rm hermit, (0)}]$ in the relaxation time approximation, from which the first-order eremitic expansion (\ref{eq:firsthermitbjor}) may be obtained.

\subsection{Two Gaussian Hot-Spots with Boost Invariance}

\begin{figure}[t]
  \includegraphics[width=0.5\linewidth]{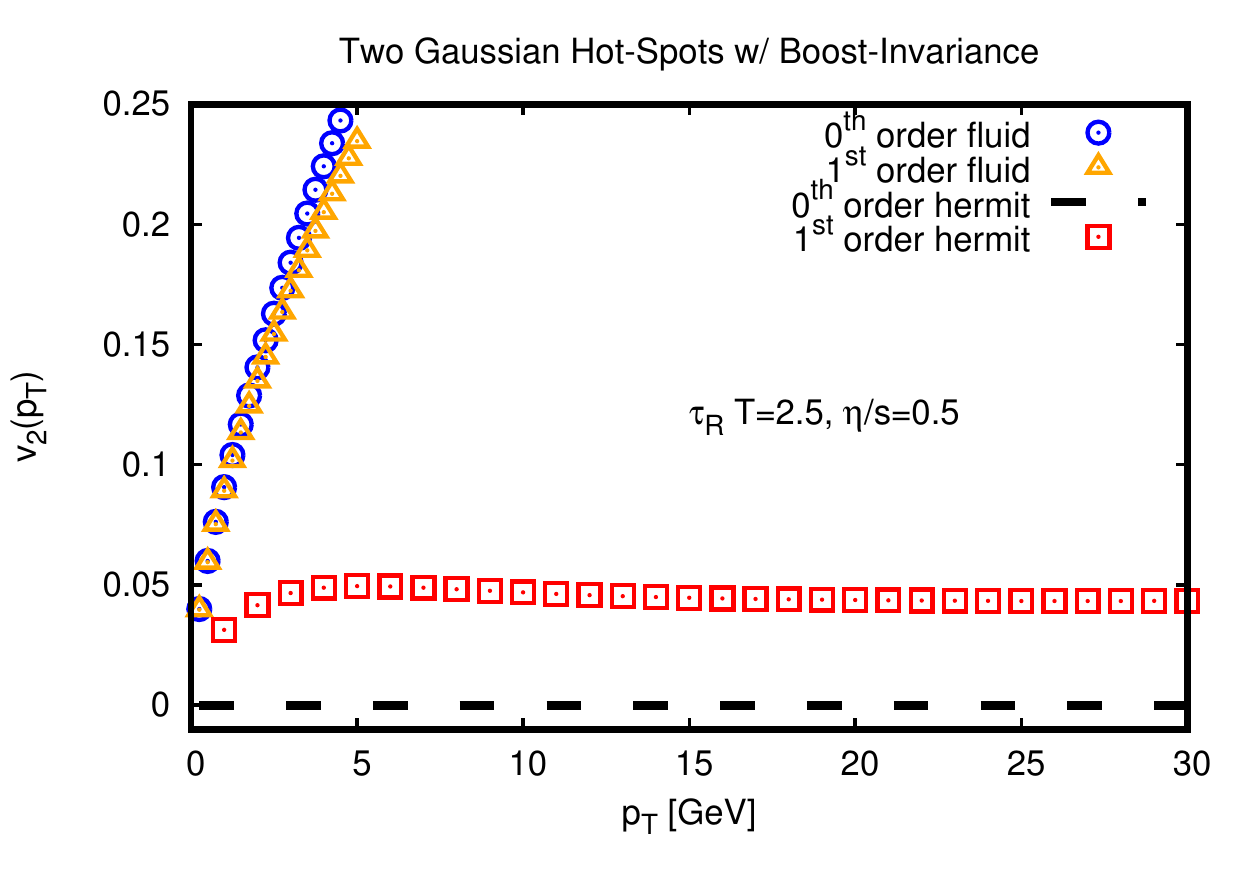}
  \caption{\label{fig:three} Parton elliptic flow $v_2(p_T)$ as a function of particle transverse momentum $p_T$ for two boost-invariant Gaussian hot-spots initially located at $x=\pm \sigma$ with $\sigma=0.4$ fm. Shown are results for zeroth and first order hydrodynamic gradient expansion as well as zeroth and first order eremitic expansion, for $\tau_R T=2.5$ (corresponding to $\frac{\eta}{s}=0.5$).}
\end{figure}

Similar to the case of Gaussian hot-spots considered in \ref{sec:two}, the superposition of two hot-spots located at position $x=\pm \sigma$ amounts to linear superposition of the contributions of individual hot-spots (\ref{eq:emtmany}) to obtain the energy-momentum tensor, from which the zeroth-order eremitic results for $\epsilon, u^\mu$, and eventually $f_{\rm hermit, (1)}$ can be obtained numerically. Of particular interest will be the so-called differential flow coefficients $v_n$ \cite{Ollitrault:1992bk,Voloshin:1994mz}, which for a boost-invariant system at mid-rapidity may be approximated as\footnote{The exact definition would imply an integration over space-time rapidity $\int d\xi$ in both numerator and denominator. However, since $f$ is strongly peaked around $\xi=Y$, the approximation $\xi=Y$ should be reasonably accurate.}
\begin{equation}
  \label{eq:vndef}
  \langle v_n(\tau,p_T)\rangle[f]\simeq \left|\frac{\int d^2{\bf x}_T \int d\phi e^{i n \phi} f(\tau,{\bf x}_T,{\bf p}_T,p^\xi=0}{\int d^2{\bf x}_T \int d\phi f(\tau,{\bf x}_T,{\bf p}_T,p^\xi=0)}\right|\,,
\end{equation}
where ${\bf p}_T=p_T\left(\cos \phi,\sin\phi\right)$.  For the case of two hot-spots with boost-invariance, the so-called elliptic flow coefficient $\langle v_2(\tau,p_T)\rangle$ may be evaluated numerically. Note that $\langle v_2(\tau,p_T)\rangle[f_{\rm hermit, (0)}]$ is time independent and vanishes identically if $\langle v_2\rangle[f_{\rm init}]=0$, consistent with the expectation that no elliptic flow is generated for non-interacting particles, while $\langle v_2(\tau, p_T)\rangle$ is in general time-dependent in first order eremitic expansion.
%\footnote{ Note that the origin of non-vanishing $\langle v_2(\tau, p_T)\rangle$ in eremitic expansions is different from the so-called escape mechanism \cite{He:2015hfa} because the latter requires emission from an anisotropic spatial surface.}. 

In order to model the flow of QCD partons, I take the number of degrees of freedom to be parametrized by
\begin{equation}
  Z=\frac{\pi^2}{15}\left(N_c^2-1+\frac{7}{4}N_c N_f\right)\,,
\end{equation}
where here and in the following the number of colors and flavors are chosen as $N_c=3$ and $N_f=2$, respectively. The initial temperature is chosen as $T_{\rm init}\simeq 0.535$ GeV at $\tau_0=0.25$ fm and the evolution is allowed to continue until a proper time $\tau_f$ at which the local temperature has dropped below $T_c=0.170$ GeV everywhere in the system (sometimes known as constant proper-time decoupling). Numerical results for $\langle v_2(\tau_f,p_T)\rangle$ are shown in Fig.~\ref{fig:three} for the first order eremitic expansion for $\tau_R T=2.5$ as a function of transverse momentum. As can be seen from this figure, the eremitic expansion suggests a near-constant behavior of the elliptic flow coefficient at large momenta. This matches the results found by Borghini and Gombeaud for the case of a two-dimensional deformed Gaussian hot-spot without boost invariance \cite{Borghini:2010hy}.

For comparison, results for $\langle v_2(\tau_f,p_T)\rangle[f_{\rm fluid}]$ using equations (\ref{eq:feq}), (\ref{eq:ff1}) are also shown in Fig.~\ref{fig:three}\footnote{The hydrodynamic results have been calculated numerically using the same initial conditions and same equation of state using the numerical solver VH2+1 \cite{Romatschke:2007mq} for the hydrodynamic equations.}.

Given that the hydrodynamic expansion breaks down at large momentum, and that the eremitic expansion breaks down at low momentum, one can expect that the ``true'' result $\langle v_2(\tau_f,p_T)\rangle[f]$ obtained from a numerical solution to the Boltzmann equation (\ref{eq:boltz}) would rise according to the hydrodynamic result at low momenta, and saturate at a constant value according to the eremitic result at high momenta for constant $\tau_R T$. It should be straightforward to test this expectation using numerical solutions to the Boltzmann equation \cite{Xu:2008av,Ruggieri:2013ova} for a two hot-spot case.

\subsection{High energy ${\rm Pb}$+$\rm Pb$ collisions}

The same techniques as for the two hot-spot case may be used to model high energy nuclear collisions, such as Pb+Pb collisions at $\sqrt{s}=5.02$ TeV. This is because the so-called Glauber model \cite{Glauber,Miller:2007ri} provides initial conditions for the matter distribution deposited after the collision as a sum over Gaussian hot-spots corresponding to the locations of collisions of the individual nucleons (see Ref.~\cite{Romatschke:2017ejr} for a recent review of relativistic nuclear collision modeling). For the purpose of this work, hot-spot locations are generated by first Monte-Carlo sampling nucleon positions for two lead nuclei from a suitably normalized Woods-Saxon probability distribution function $\rho({\bf x})\propto (1+e^{(|{\bf x}|-r_0)/a_0})^{-1}$ with $r_0=6.62$ fm and $a_0=0.546$ fm. Random sampling of impact parameters for the collision of two lead nuclei, nucleons are said to undergo a collision if their respective distance in the transverse ${\bf x}_T$ plane is less than $|{\bf x}_T|<\sqrt{\sigma_{NN}/\pi}$, where $\sigma_{NN}\simeq 60$ mb is the (collision-energy dependent) nucleon-nucleon cross-section at $\sqrt{s}= 5.02$ TeV. Each location of a collision is taken to correspond to the location of one Gaussian hot-spot. The sum over these Gaussian hot-spots defines the initial energy-density distribution in the transverse plane, which is successfully used in modern hydrodynamic modeling of lead-lead collisions \cite{Romatschke:2017ejr}. The number of nucleons participating in a collision is related to the total entropy of the system, which in turn translates to the number of particles observed in experiment (``multiplicity''). In the following, I will consider mid-central lead-lead collisions corresponding to the 30-40 percent highest multiplicity class. For this case, the parameter $T_{\rm init}$ was adjusted such that the hydrodynamic evolution with a QCD equation of state \cite{Borsanyi:2010cj} gives multiplicities that are consistent with those found in experiment \cite{Adam:2015ptt}. Unlike the two hot-spot case treated above, for multiple hot-spots found in the Glauber modeling of Pb+Pb collisions the momentum flow coefficients $v_n$ for $n=3,4,\ldots$ are also non-vanishing in general.

Last but not least, the relaxation time coefficient for QCD is expected to scale as \cite{Hosoya:1983xm,Arnold:2003zc}
\begin{equation}
  \label{eq:taurT}
  \tau_R \propto T^{-1} \alpha_s^{-2}(Q^2)\,,
\end{equation}
up to logarithmic corrections, where $\alpha_s(Q^2)$ is the QCD coupling that in one-loop running is given by 
\begin{equation}
  \alpha_s(Q^2)=\frac{4 \pi}{(11-\frac{2}{3}N_f) \ln Q^2/\bar{\Lambda}^2}\,,
\end{equation}
where $\bar{\Lambda}\simeq 0.376$ GeV to match the experimentally determined value $\alpha_s(Q^2=M_z^2)=0.1184$ at the mass of the Z-boson $M_z\simeq 91.18$ GeV \cite{Olive:2016xmw}. I therefore use $\tau_R T=\frac{1}{7}\alpha_s^{-2}(p_T^2)$ when attempting to make comparisons to QCD (numerically, this implies $\eta/s=\frac{\tau_R T}{5}\simeq 0.13$ when evaluating $Q=p_T\simeq 1.5$ GeV). It should be emphasized that the choice $\tau_R T \alpha_s^2=\frac{1}{7}$ is arbitrary, and has been made for illustration purposes. Nevertheless, in view of the sizable uncertainty in the value of e.g. $\eta/s$ from perturbative QCD calculations at scales $p_T\simeq 1.5$ GeV, such a choice does not seem entirely unreasonable \cite{Ghiglieri:2018dib}.

\begin{figure}[t]
  \includegraphics[width=0.45\linewidth]{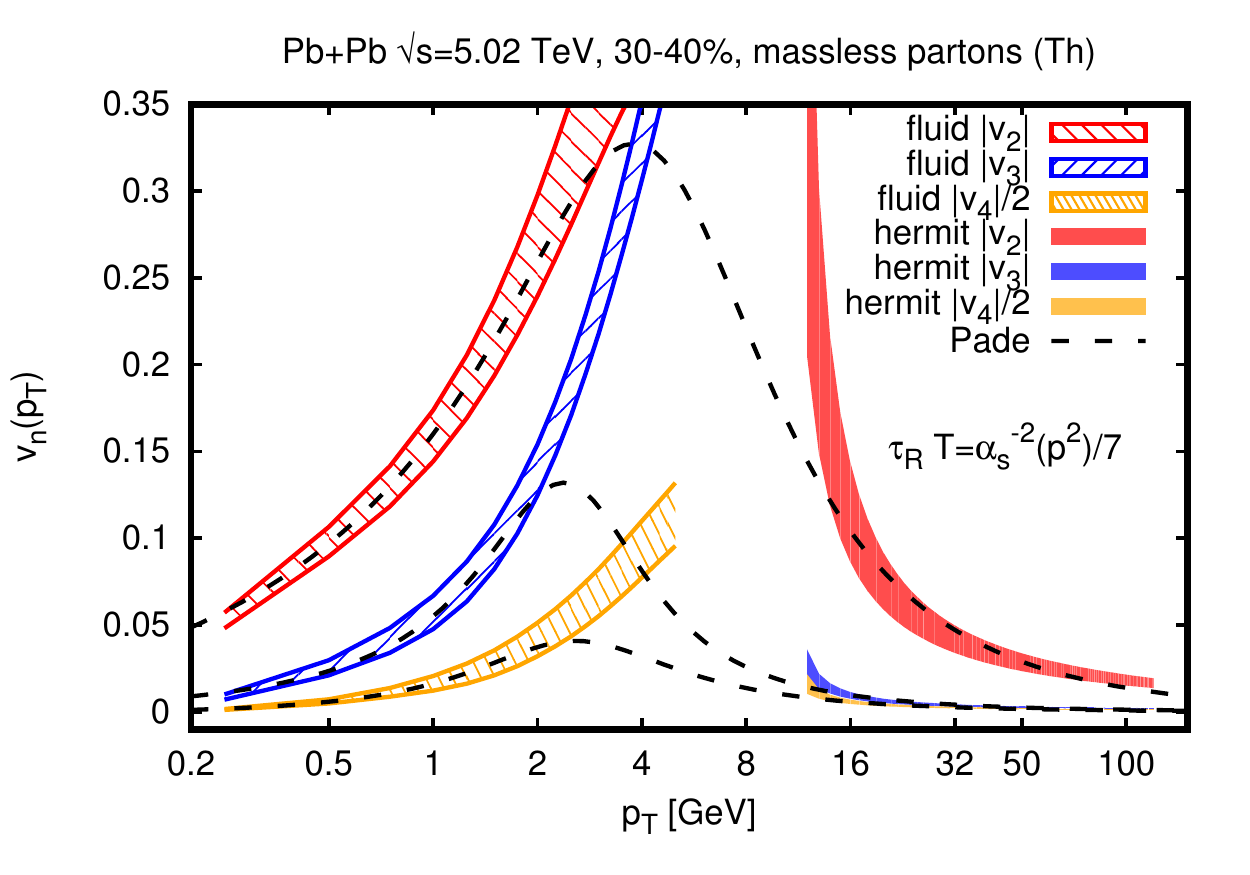}
  \hfill
  \includegraphics[width=0.45\linewidth]{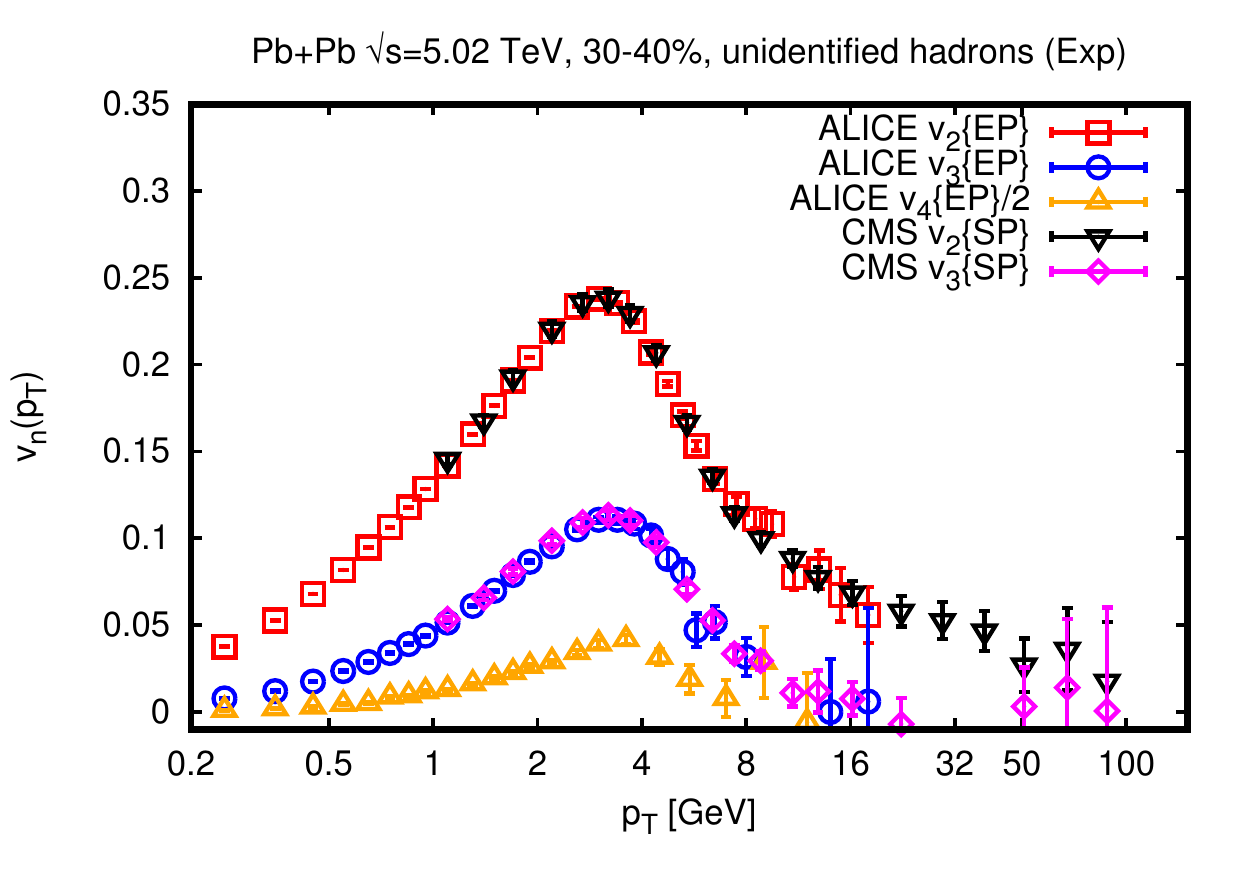}
  \caption{\label{fig:four} Left: Results for momentum anisotropy coefficients $\langle v_n(p_T)\rangle$ for massless partons from theoretical calculations at low momenta (zeroth order hydrodynamic gradient expansion) and high momenta (first order eremitic expansion). For illustration, low and high momentum results are connected through Pad\'e-type fits. Right panel: experimental data \cite{Abelev:2012di,Chatrchyan:2012xq,Sirunyan:2017pan} for momentum anisotropy coefficients for unidentified hadrons in Pb+Pb collisions at $\sqrt{s}=5.02$ TeV in the 30-40 percent multiplicity class.}
\end{figure}

Results for the anisotropic flow coefficients $\langle v_n (\tau_f,p_T)\rangle$, averaged over 10 events of initial hot-spot locations in the 30-40 percent multiplicity class of Pb+Pb collisions are shown in Fig.~\ref{fig:four}. Again, results from zeroth order eremitic expansion for $\langle v_n(\tau_f,p_T)\rangle$ vanish identically, while first order results differ significantly from zero for $\langle v_2\rangle [f_{\rm hermit, (1)}]$. For $p_T\lesssim 15$ GeV, the eremitic expansion seems to break down for this choice of $\tau_R$ since the correction $f_{\rm hermit, (1)}$ approaches the leading-order result $f_{\rm hermit, (0)}$. At high $p_T$, $\langle v_n(\tau_f,p_T)\rangle$ appears to approach a constant times the inverse of $\tau_R T$, just as what was found in the two hot-spot case above. For $\tau_R T$ of the form (\ref{eq:taurT}), this implies $\langle v_n (\tau_f,p_T)\rangle$ falling as the QCD coupling constant squared for large $p_T$. For comparison, results for $\langle v_n(\tau_f,p_T)\rangle[f_{\rm fluid, (0)}]$ are also shown in Fig.~\ref{fig:four}. The hydrodynamic gradient expansion breaks down at high $p_T$, but $\langle v_n(\tau_f,p_T)\rangle[f_{\rm fluid, (0)}]$ exhibits the rising trend familiar from full hydrodynamic simulation studies \cite{Romatschke:2017vte}.
The hydrodynamic gradient results\footnote{Note that the hydrodynamic curves shown in Fig.~\ref{fig:four} were calculated with a QCD equation of state \cite{Borsanyi:2010cj} instead of an ideal equation of state to increase numerical stability of the hydrodynamic solver.}  at low momenta $p_T\leq 2$ GeV can be connected to the eremitic curves at high momenta $p_T\geq 15$ GeV by a type of Pad\'e fit, suggesting a peak in $\langle v_n(\tau_f,p_T)\rangle$ for specific values of $p_T$ for $n=2,3,4$. Note that the available information at low and high momenta, respectively, is not sufficient to unambiguously determine the location or height of the peaks in $\langle v_n(\tau_f,p_T)\rangle$.

Since the results shown for $\langle v_n(\tau_f,p_T)\rangle$ are for massless partons obtained when the whole system has cooled down below a pre-defined temperature, the results are not directly comparable to experimental data. However, it is tempting to inspect the relevant experimental data on differential flow coefficients for 30-40\% Pb+Pb collisions for unidentified hadrons, shown in the rhs panel of Fig.~\ref{fig:four}. Interestingly, the experimental data seems to exhibit the qualitative features of the above theoretical calculations at low momenta (rise with $p_T$ as predicted by hydrodynamic expansions) and high momenta (decrease with $p_T$ as predicted by eremitic expansions). Curiously, also the magnitude of experimentally measured $v_n$ coefficients at $p_T\lesssim 2$ GeV and $p_T \gtrsim 15$ GeV seem to be consistent with theoretical calculations shown in the lhs panel of Fig.~\ref{fig:four}. Furthermore, note that the ratio $\frac{\langle v_3(\tau_f,p_\perp)\rangle}{(\langle v_2(\tau_f,p_\perp)\rangle)^{3/2}}\simeq 1$ exhibits near-constant behavior close to unity as a funtion of $p_T$ at large momenta in eremitic expansion, similar to what has been observed experimentally \cite{ATLAS:2012at}.

\section{High energy $p$+$\rm Pb$ collisions}

\begin{figure}[t]
  \includegraphics[width=0.45\linewidth]{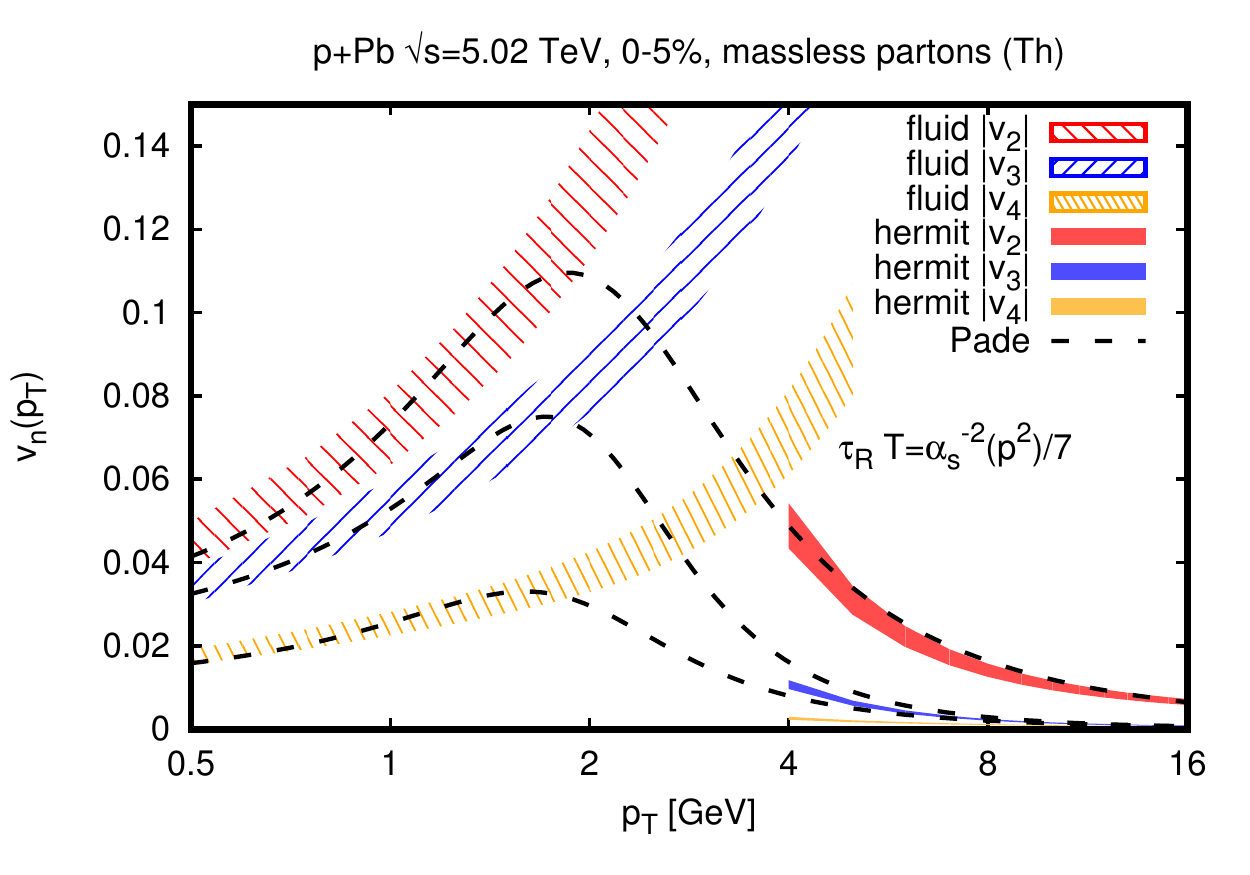}
  \hfill
  \includegraphics[width=0.45\linewidth]{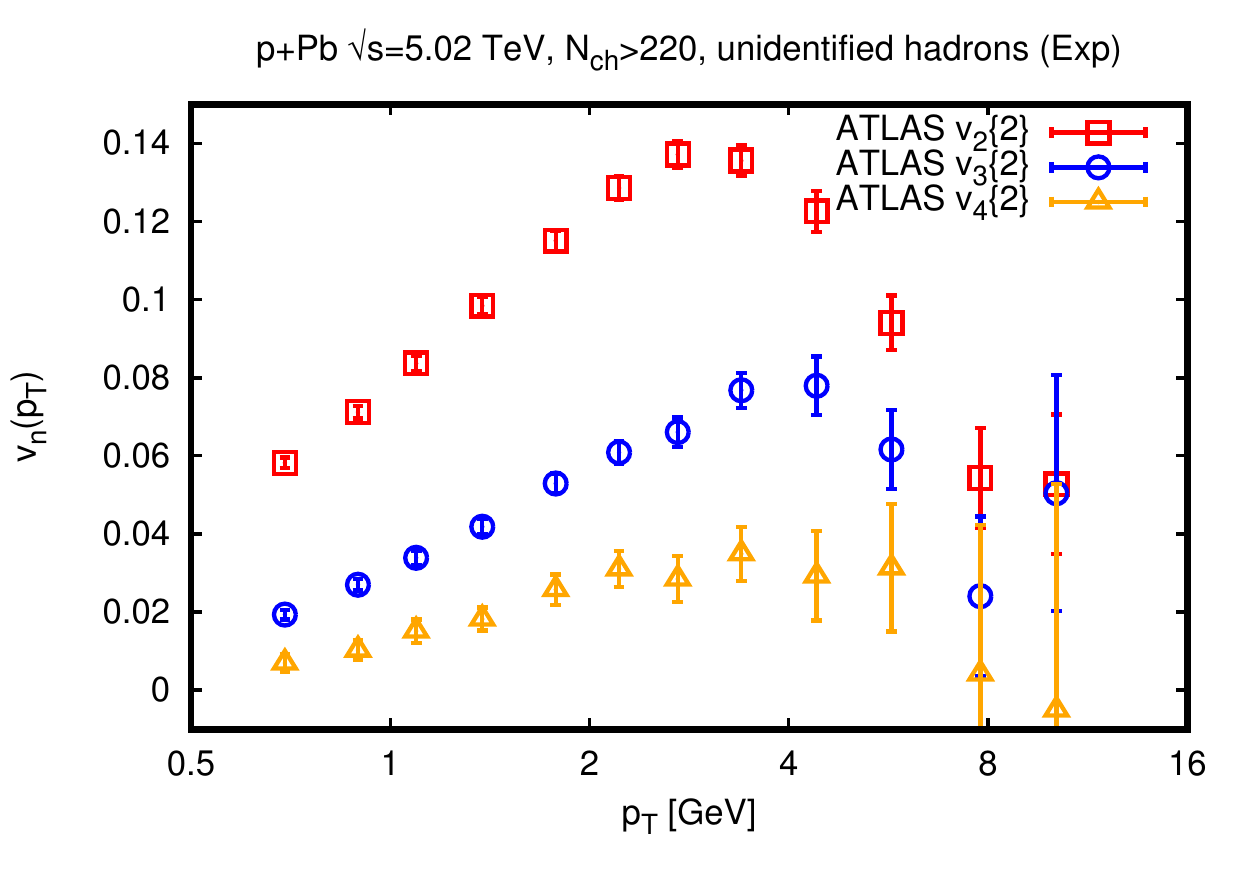}
  \caption{\label{fig:five} Left: Results for momentum anisotropy coefficients $\langle v_n(\tau_f,p_T)\rangle$ for massless partons from theoretical calculations at low momenta (zeroth order hydrodynamic gradient expansion) and high momenta (first order eremitic expansion). For illustration, low and high momentum results are connected through Pad\'e-type fits. Right panel: experimental data \cite{Aad:2014lta} for momentum anisotropy coefficients for unidentified hadrons in p+Pb collisions at $\sqrt{s}=5.02$ TeV for central collisions.}
\end{figure}

One of the unresolved questions in the context of high energy nuclear collision is the mechanism for the measured sizable $v_2$ coefficient at high transverse momenta $p_T\gtrsim 10$ GeV, cf. Fig.~\ref{fig:four}. It has been suggested that the measured $v_2$ coefficient arises from jet quenching, with highly energetic particles (jets) losing more energy when traveling through a longer path length in a medium \cite{Zhang:2013oca,Betz:2016ayq}. However, jet quenching seems to be absent in proton-lead collisions, yet the experimentally measured $v_2$ coefficient exhibits the same behavior as in lead-lead collisions \cite{Nagle:2018nvi}, cf. Fig.~\ref{fig:five}. Eremitic expansions offer a potential alternative explanation for the observed $v_2$ coefficient, namely through non-hydrodynamic transport of the initial geometry. While the momentum anisotropies in eremitic expansions arise from the dynamics of high energy particles, these particles are nevertheless part of, and flowing with, the medium, as opposed to the modeling of jets, which are by definition treated separately from the medium.

For this reason, I have simulated central p+Pb collisions through Monte-Carlo sampling positions of nucleon collisions from a Glauber model, and using these positions as the initial location of Gaussian hot-spots as explained in the preceding sections. The dynamics encountered in p+Pb is not boost-invariant, but hydrodynamic simulations seem to indicate that nevertheless boost-invariance is not a bad quantitative approximation in practice \cite{Schenke:2014zha,Romatschke:2015gxa,Bozek:2015swa,Weller:2017tsr}. The results for the momentum anisotropies $\langle v_n(\tau_f,p_T)\rangle$ averaged over 10 events for zeroth order hydrodynamic and first order eremitic expansion are shown in Fig.~\ref{fig:five}. One finds that the same qualitative features as in Pb+Pb emerge: rising $v_n$ coefficients at low $p_T$ as predicted by hydrodynamics, and falling $v_n$ coefficients at large $p_T$ as predicted by eremitic expansions. Unlike the case for Pb+Pb collisions, the magnitude for $\langle v_2\rangle$ at large $p_T$ for massless partons from eremitic expansions of p+Pb collisions, while non-vanishing, is systematically below the experimentally measured values for unidentified hadrons (rhs panel of Fig.~\ref{fig:five}). Future studies involving more realistic equations of state and a confinement prescription will be needed in order to decide if eremitic expansion qualify as explanation for the observed $v_2$ coefficient at high momenta.

\section{Summary and Conclusions}

In this work, a systematic expansion procedure for relativistic kinetic theory in the large mean-free path regime was considered. This eremitic expansion procedure is complementary to the perhaps more familiar hydrodynamic expansion scheme in that it allows controlled calculations at very high particle momenta, while breaking down at low particle momenta. Eremitic expansions allow to probe purely non-hydrodynamic transport phenomena since hydrodynamic modes are absent in this approach. Using kinetic theory in the relaxation time approximation as an example, first order eremitic expansions for Gaussian hot-spots with and without boost invariance were calculated. Applications for these calculations to evaluating the momentum anisotropy coefficients $v_n(\tau_f,p_T)$ in Pb+Pb and p+Pb collisions at $\sqrt{s}=5.02$ TeV were presented, and it was found that eremitic expansions qualitatively describe the experimentally measured behavior of flow coefficients at high momenta. Thus, eremitic expansions offer a potential alternative to jet quenching as the source for the measured elliptic anisotropy at high momenta.

Many generalizations and validations of the present work are possible. For instance, the second order correction to eremitic expansions should be straightforward to calculate for many of the examples given in this work. The quantitative reliability of eremitic expansions should be checked by direct comparison to full numerical solutions of the Boltzmann equation. The application of eremitic expansions to relativistic collision systems should be made more realistic by including a QCD equation of state and a hadronization procedure.

Nevertheless, eremitic expansions seem to have the potential to become an interesting new tool in the study of relativistic collision systems and the phenomenology of non-hydrodynamic transport.

\section{Acknowledgments}

This work was supported in part by the Department of Energy, DOE award No DE-SC0017905. I would like to thank Jamie Nagle, Nicolas Borghini, Ulrike Romatschke and J\"urgen Schukraft for helpful comments, and Gowri Sundaresan for collaboration in the early stages of this work.

\bibliographystyle{hunsrt}
\bibliography{pp-hydro}

\end{document}